# Representing caregiver burden in observational studies: Development of the Caregiver Burden Index (CareBI) using NSOC

[a]Forough Mahpouya, [a]Sabrina Casucci, [b]Suzanne Sullivan, [c]Christopher Barrick

## Abstract

**Background and Objectives:** Informal caregiving often carries a significant emotional, physical, and financial toll, yet caregiver burden is often underrepresented in healthcare research and methods. Existing caregiver burden instruments, while valuable in clinical research, often lack compatibility with observational datasets regularly used in health services research and planning. This study introduces the Caregiver Burden Index (CareBI) developed for the National Study of Caregiving (NSOC), that can be used to represent caregiver burden in quantitative models and observational research studies.

**Research Design and Methods:** CareBI was developed and validated using a multistep process that included the identification and preparation of individual NSOC survey items, exploratory and confirmatory factor analysis, score estimation, interpretation, and external validation. The study used data from round 12 of the NSOC.

**Results:** CareBI represents three domains of burden: objective, subjective, and interpersonal, providing a comprehensive view of both the positive and negative aspects of caregiving. It also aligns with the Zarit Burden Interview, a widely used tool for prospectively assessing caregiver burden. Construct validity was assessed by comparing CareBI's relationship with caregiver and care recipient outcomes, as well as sensitivity to known burden-related risk and mitigation factors. Early findings affirm the scale's utility in categorizing low-, moderate-, and high-burden caregivers and guiding resource-oriented strategies.

**Discussion and Implications:** CareBI represents a reproducible tool for embedding caregiver metrics into health operations, predictive modeling, and public policy frameworks, and provides a template for applying operations research and industrial engineering methods to psychosocial measurement challenges in aging and long-term care.

[a] University at Buffalo, Department of Industrial and Systems Engineering. Forough Mahpouya: foroughm@buffalo.edu, Sabrina Casucci: scasucci@buffalo.edu

[b] SUNY Upstate Medical University, College of Nursing. Suzanne Sullivan: SullivSu@upstate.edu

[c] University at Buffalo, School of Nursing. Christopher Barrick: cbarrick@buffalo.edu

## Background and Objectives

The population of older adults in the United States is growing rapidly. Projections indicate that by 2060, one in four Americans will be 65 years of age and older (Vespa et al., 2018). Family caregivers are essential to supporting older adults who wish to "age in place" or remain in their homes for as long as possible (Robinson-Lane et al., 2023; Wolff et al., 2025). In 2022, the number of family caregivers in the United States increased to 24.1 million, a nearly 32% increase over from 2011 (Wolff et al., 2025). Informal caregivers are typically unpaid yet play an increasingly important role over time in helping with daily activities and medical tasks (Kim et al., 2023). There are, however, few supports for this critical population. Access to respite services remains inadequate. Caregiver training opportunities are sparse and many caregivers lack access to financial assistance (Green et al., 2024), which can exacerbate caregiving difficulties and impacts. While informal caregiving can produce positive effects for the care recipients and caregivers, the negative impacts on the caregivers' physical, mental, and emotional well-being can have serious and lasting effects, including care recipients' premature transition to institutional care or death, as well as caregivers' lost income and reduced quality of life (Sullivan et al., 2023; Marinacci, L. X., 2025).

Accurately measuring caregiver burden is therefore critical for understanding the scope of this public health issue, enabling accurate identification of at-risk individuals, and developing more effective interventions to lessen the negative impacts of caregiving (Griffin, J. M., 2024). Assessment tools, such as the Zarit Burden Interview (ZBI) (Zarit et al., 1987), Caregiver Strain Index (CSI) (Robinson, 1983), and Caregiver Burden Scale (CB scale) (Elmstahl et al., 1996), are effective means for assessing burden prospectively, making them valuable tools in clinical care. Yet, health informatics researchers lack equivalent measures and validated approaches when using observational data, limiting our ability to derive caregiving related insights from this essential data source.

There are several challenges in using observational data to develop a scale (here, a measure of caregiver burden). First, traditional methods, developed for prospectively administered surveys must be modified to accommodate observational data (DeVellis and Thorpe, 2021). Second, scale development using observational studies is inherently limited to the information available in the dataset. The thoroughness and quality of the observational data set is therefore paramount to ensure the presence of all necessary items (i.e., questions in a survey) that can accurately represent the burden of caregiving for older adults. Further, as is the case with many surveys or questionnaires, the response approach to the relevant items may differ, making it difficult to combine them into a single summary scale; for example, one item might be answered in binary form (e.g., Yes/No), whereas another might involve a multi-level Likert-type response. Finally, the lack of ground truth in the dataset—i.e., the true caregiver burden, such as a single question asking directly about burden or an established measurement like the ZBI—complicates testing the validity of the proposed scale.

The 22-item ZBI tool, in widely used in interviews and self-assessment, that represents the physical, emotional, social, and financial aspects of perceived caregiver burden on a scale of 0-88 (Al-Rawashdeh et al., 2016). Critically, all ZBI items were designed to use the same 5-level Likert-type responses, from 0 ("never") to 4 ("nearly always"), indicating how often the caregiver feels burdened in a specific aspect of caregiving. While this approach facilitates prospective data collection and burden assessment, observational data often combines multiple data types and inconsistent response scales, limiting its use in observational studies.

For example, Riffin et al. (2019) created a scale from National Study of Caregiving (NSOC) items on emotional, physical, and financial caregiving challenges, with each question having a binary (Yes/No) response style. The scale reported participants as those with no burden and those with any burden, without assessing the severity of the burden. Halpern et al. (2017) measured the burden on caregivers of cancer patients using the NSOC and NHATS datasets but separately assessed emotional burden, psychological burden, and the caregiver's relationship with the patient. Spillman et al. (2014), grouped NSOC questions into three categories: caregiving gains, negative aspects of caregiving, and caregiving difficulties, highlighting the range of items available in NSOC. Later, Freedman and Wolff (2025) created the NSOC Caregiver Strain Scale (NSOC-CSS) to measure both emotional and non-emotional strain, adapting components from the established ZBI-22 and Robinson's (1983) CSI frameworks to measure the negative aspects of caregiving. While each of these examples provides critical insights into the domains of caregiving burden, there remains a critical need for a quantitative representation, or scale, that comprehensively reflects both the multi-dimensional nature of burden and its severity to facilitate quantitative observational research.

This study aims to fill this gap by developing a multidomain Caregiver Burden Index (CareBI) that synthesizes numerous factors of caregiver burden related to caring for older adults into a single value where a higher value of this metric represents a greater severity of burden. Our method develops a single value representing caregiver burden (CareBI) from observational data by adapting traditional scale development approaches of DeVellis and Thorpe (2021), which were originally designed for prospective scale development. To facilitate the use of CareBI in clinical research and care, we propose an interpretation approach that categorizes stages of burden. Finally, we evaluate the reliability and validity of the developed scale.

## Research Design and Methods

**Dataset**

**NHATS/NSOC.** The NSOC is a longitudinal, nationally representative survey of 2007 caregivers of 1369 Medicare beneficiaries who responded to the National Health and Aging Trends Study (NHATS). The NSOC includes caregiver characteristics, caregiving contexts, and associated outcomes, which makes it a valuable resource for examining caregiver burden (Eden and Schulz, 2016). The NHATS is sponsored by the National Institute on Aging (grant number NIA U01AG32947) and was conducted by the Johns Hopkins University. For this study, we used data from round 12 (2022) with permission under a Data Use Agreement with the administrators of NHATS/NSOC. The NSOC is the most current observational dataset on caregivers available at the time of this paper.

**Inclusion Criteria and Definitions**

**Caregivers.** Caregivers were defined as any family or other unpaid caregivers who assisted NHATS participants with self-care, mobility, or household activities within the past month or year. Caregivers of NHATS participants living in nursing homes, those with only a Facility Questionnaire, or those who were deceased at the time of interview were excluded. Caregivers who did not provide help during the month prior to the interview were excluded to ensure the study reflected recent, active experiences.

**Caregiver burden.** We defined caregiving burden as the caregiver's perception of distress resulting from the demands of providing care. We began by reviewing the key aspects of burden, including objective burden (i.e., actual demands of the care situation), and subjective burden (i.e., caregiver's personal view of being stressed) (Siminoff, L. A., 2024). Our work aligns with Zarit et al. (1987), who determined that a caregiver burden scale must capture how caregivers experience the negative effects of caregiving, and their impact on emotional, social, financial, physical, and spiritual well-being and Iecovich (2011), who concluded that a better quality of the relationship between the caregiver and care recipient is associated with lower caregiver burden. Therefore, the items selected for CareBI from the NSOC dataset were based on conceptual alignment with caregiving domains rather than a direct item-by-item mapping to a well-known scale.

**Scale Development Procedure**

We used an iterative process shaped by both statistical analysis and conceptual interpretation (see Figure 1). Missing values of less than 10% for the selected items were imputed with the most frequent responses (Alwateer et al., 2024), resulting in 2,155 observations for scale development. We reverse-coded responses that were not coded in ascending order, to ensure that higher numerical values reflected greater burden (see Supplementary Figure 1). All analyses were conducted in RStudio 2024.12.0+467.

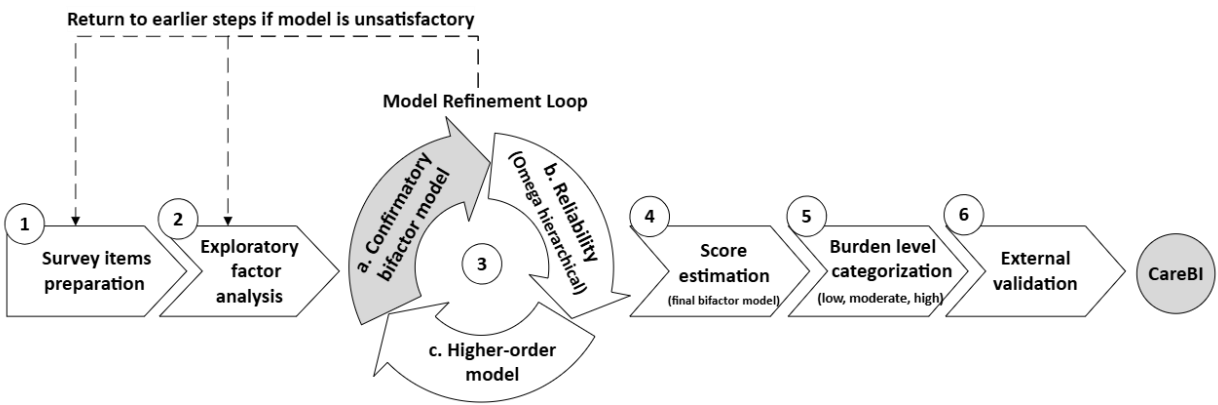

**Figure 1**. Overview of the CareBI scale development process.

**Step 1. Survey Item Preparation.** We began with a pool of items in the NSOC dataset that reflect the burden caregivers feel or perceive when providing care to an older adult, including emotional and physical stress to the burden in their private and social lives. Items were selected that directly reference "helping the care recipient" as well as questions that incorporate emotional stress but do not explicitly reference "helping the care recipient." We relied on the principle of attribution theory in our selection, assuming that when an individual's life is dominated by a salient role, such as that of a caregiver, they tend to attribute their overall emotional and psychological states to that central experience (Kelley and Michela, 1980). So, even a general question like "How often did you feel bored?" is likely answered through the lens of the caregiving context where caregivers implicitly link their feelings of boredom to the duties of providing care.

We then developed new composite items by combining pairs of related questions into a single, multi-level ordinal item. For instance, the NSOC question "Is helping care recipient emotionally difficult for you?" (Yes/No) and "How emotionally difficult is it helping the care recipient?" (1. a little difficult–5. very difficult) were merged into a single composite item where caregivers who responded "No" were assigned the minimum burden score (e.g., 0) and caregivers who responded "Yes" were assigned a value equal to their responses ranging from 1 to 5.

When necessary, we transposed, or reverse-coded, question responses to ensure that higher numerical values represented greater perceptions of burden. Supplementary Figure 1 provides an example of this transformation.

**Step 2. Exploratory Factor Analysis.** We first used the Kaiser-Meyer-Olkin (KMO) measure of sampling adequacy to ensure that selected items were suitable for factor analysis, including only those items where KMO ≥ 0.80 (Kaiser, 1974). We then determined the number of unrotated common factors using parallel analysis (Horn, 1965) basing our analysis on the polychoric correlation matrix, which can accommodate different response formats (Yes/No, Likert) (Revelle, 2025). Decisions on variable inclusion were further supported by an examination of the scree plot, which visually indicates the optimal number of factors at the "elbow" (point of inflection) in the graph of successive eigenvalues (Zhu and Ghodsi, 2006).

Exploratory factor analysis was completed using only those factors suggested by parallel analysis. Because the item distributions were skewed, we estimated the factor model with the minimum residual (minres) method, which is a robust technique when normality assumptions are violated (Revelle, 2025). To simplify the factor structure and improve interpretability, we rotated the results using direct oblimin—an oblique rotation method chosen as the factors are assumed to be correlated, as expected in our model (Corner, 2009).

To achieve a simple structure, i.e., each item demonstrates a clear, meaningful, and strong loading on a single factor, we retained items based on three statistical criteria: (1) a primary factor loading exceeding 0.40 (Cheung et al., 2024), (2) no significant cross-loadings (i.e., loadings below 0.32 on all other factors) (Costello and Osborne, 2005), and (3) a communality value of at least 0.40 (Costello and Osborne, 2005). We consistently examined the factors to ensure that all items loading on a factor conceptually convey the same understandable concept, which allowed for the retention of conceptually informative items that refined the overall framework, even when they moderately violated statistical cutoffs.

Finally, we evaluated the goodness-of-fit of alternative factor models using three indices: (1) the Tucker Lewis Index (TLI), which treats values of 0.95 or higher as evidence of excellent fit (Marsh et al., 2004), (2) the root mean square error of approximation (RMSEA) with values of 0.06 or below indicating a close fit with the data (Marsh et al., 2004), and (3) the Bayesian information criterion (BIC) which considers model parsimony, with lower numbers signaling a better balance between model fit and complexity (Kuha, 2004).

**Step 3. Model Refinement Loop.** We developed CareBI using a bifactor model. We used a model refinement loop (see Step 3 in Figure 1) to establish a final structure for caregiver burden that was both robust (to nonnormal, ordinal, and categorical data) and interpretable. Our goal was to partition item variance, allowing an examination of how much variance is explained by the general factor versus the specific domains (Reise et al., 2010). This iterative process began by specifying the EFA-derived structure as a confirmatory bifactor model (step 3a). In this model, each item loads on a single general factor representing the target "burden" construct, and on at most, one orthogonal group factor that captures additional variance from specific items in the group (caregiving burden domain). We estimated the confirmatory bifactor model using the weighted least squares mean and variance adjusted (WLSMV) estimator as it is designed for analyzing ordinal or categorical data, such as Likert-scale responses (Li, 2016).

After fitting, we evaluated the general factor score reliability (step 3b) using coefficient omega hierarchical ($\omega_h$), with a value of 0.70 or higher considered to show acceptable reliability (Reise et al., 2010; Forbes et al., 2021). We then fit a corresponding higher-order factor model at each stage (Step 3c) to clarify the hierarchical links among latent domains, and to guide interpretation and revisions of the bifactor model. The fitting, interpreting, and refining loop (step 3.a to 3.c) was repeated until the bifactor model showed strong statistical fit, exemplified through three criteria: (1) TLI ≥ 0.9, (2) comparative fit index (CFI) ≥ 0.9 (Bentler, 1990), and (3) RMSEA ≤ 0.08. The final model was also required to meet acceptable reliability ($\omega_h \geq 0.70$) and to have a clear conceptual meaning.

We used the bifactor model as our final scoring model because it separates the general caregiver burden factor from domain-specific variance (group factors) and provides reliable individual scores. A higher-order model was only fit to support conceptual interpretation of the domains.

**Step 4. Score Estimation.** CareBI score estimates were derived from the finalized bifactor model. First, individual scores for the latent general factor, "burden", were estimated for each participant using a regression-based method, as implemented in the lavaan package in R (Thurstone, 1934; Rosseel, 2012). This procedure produces continuous factor scores that are on a z-score-like scale and represent the score of each individual on the pure general factor (a latent score purified from domain-specific variances and errors).

To enhance interpretability, the raw factor scores were then transformed onto a 0–100 scale using a standard min-max normalization technique where the lowest observed general factor score was mapped to 0 and the highest observed score was mapped to 100 (Han et al., 2006). Finally, the rescaled scores were rounded to the nearest integer to produce the final CareBI score.

**Step 5. Burden Level Categorization.** To translate the final continuous CareBI scores (scaled 0-100) into meaningful, practical categories, such as low, moderate, and high burden, we used the data-driven k-means clustering approach (MacQueen, 1967) to discover if natural groupings existed within the scores themselves. This algorithm finds the centers of the densest "clumps" of scores (a group of scores that are closest to each other). We searched for three such groups (k=3) and set the halfway points between the centers of these three empirically identified groups as the thresholds for score categorization.

**Step 6. External Validation.** We assessed CareBI's validity using criterion and construct validation procedures (Boateng et al., 2018). Mixed-effects regression analysis was conducted to account for the non-independence of data, because multiple caregivers are often nested within the same care recipient in the NSOC dataset, and their shared circumstances could influence their burden scores (Brooks et al., 2017).

To demonstrate criterion validity, we evaluated CareBI's ability to predict theoretically related outcomes, testing whether the final CareBI scores significantly predicted subsequent adverse outcomes, including caregiver-related outcomes such as mental health deterioration (e.g., feeling depressed, anxious, or hopeless) and poor sleep, as well as care recipients' related outcomes (e.g., having an overnight hospital stay).

Construct validity was assessed by exploring both convergent and discriminant aspects. The convergent validity tested whether the CareBI score was positively associated with factors known to increase caregiver burden, such as the intensity of care (e.g., hours spent helping with daily activities) and also the range of caregiving responsibilities (e.g., managing medications, shopping for groceries, and assisting with banking). In contrast, discriminant validity tested whether the CareBI score was, as would be expected, negatively associated with factors known to mitigate burden, such as receiving help from family and friends or using services that provide respite care, confirming that the scale is distinct from concepts of caregiver self-efficacy or support.

If the model did not meet predefined criteria for statistical fit, reliability, or conceptual clarity, we returned back to the item selection or EFA for further refinement until a satisfactory model was achieved. Practical steps to measure CareBI are provided in Supplementary Table 3.

## Results

### Scale Measurement Results

An initial pool of 44 items from the NSOC survey was selected for conceptual alignment with the domains of the 22 question ZBI (Supplementary Table 1). The KMO measure of 0.92 confirmed the factorability of this initial item set. Using the iterative scale development process and confirmatory model testing, we retained a set of 18 items in our final model (Figure 2). For this 18-item set, the KMO measure remained high at 0.87, and parallel analysis suggested a six-factor solution as the optimum number of factors. The final EFA model fitted the data well (TLI = 0.952, RMSEA = 0.057, and BIC = 12.32), demonstrating six clear domains (steps 1 and 2): overload, difficulty, mood, health, social participation, and relationship quality.

Overload captured the feeling of being overwhelmed by caregiving demands and the lack of personal time, while Difficulty reflected the physical, financial, and emotional challenges associated with caregiving. Mood and Health factors demonstrated the caregivers' internal stress on their emotional state and physical well-being, respectively. Social participation captured the caregiver's ability to engage in a personal social life, and Relationship quality measured the nature of the interaction with the person they are caring for.

We used a confirmatory bifactor model as the final estimation model for CareBI. In this framework, every item loads on a general burden factor, the primary construct of interest, and at most on one orthogonal group factor that represents domain specific variance. The bifactor model showed good overall fit (robust CFI = 0.959, robust TLI = 0.946, robust RMSEA = 0.060) and provided a direct way to measure burden by producing a purified general factor score based on all 18 items. The general factor demonstrated acceptable reliability ($\omega_h$ = 0.74) and explained most of the shared variance. In contrast, the reliabilities of the group factors ($\omega_s$), once the general factor was accounted for, were quite low: overload (0.28), difficulty (0.15), mood (0.49), health (0.29), social participation (0.30), and relationship quality (0.41). Because the general factor was strong and the group factors were weak, we report a single CareBI score rather than using group scores as independent outcomes.

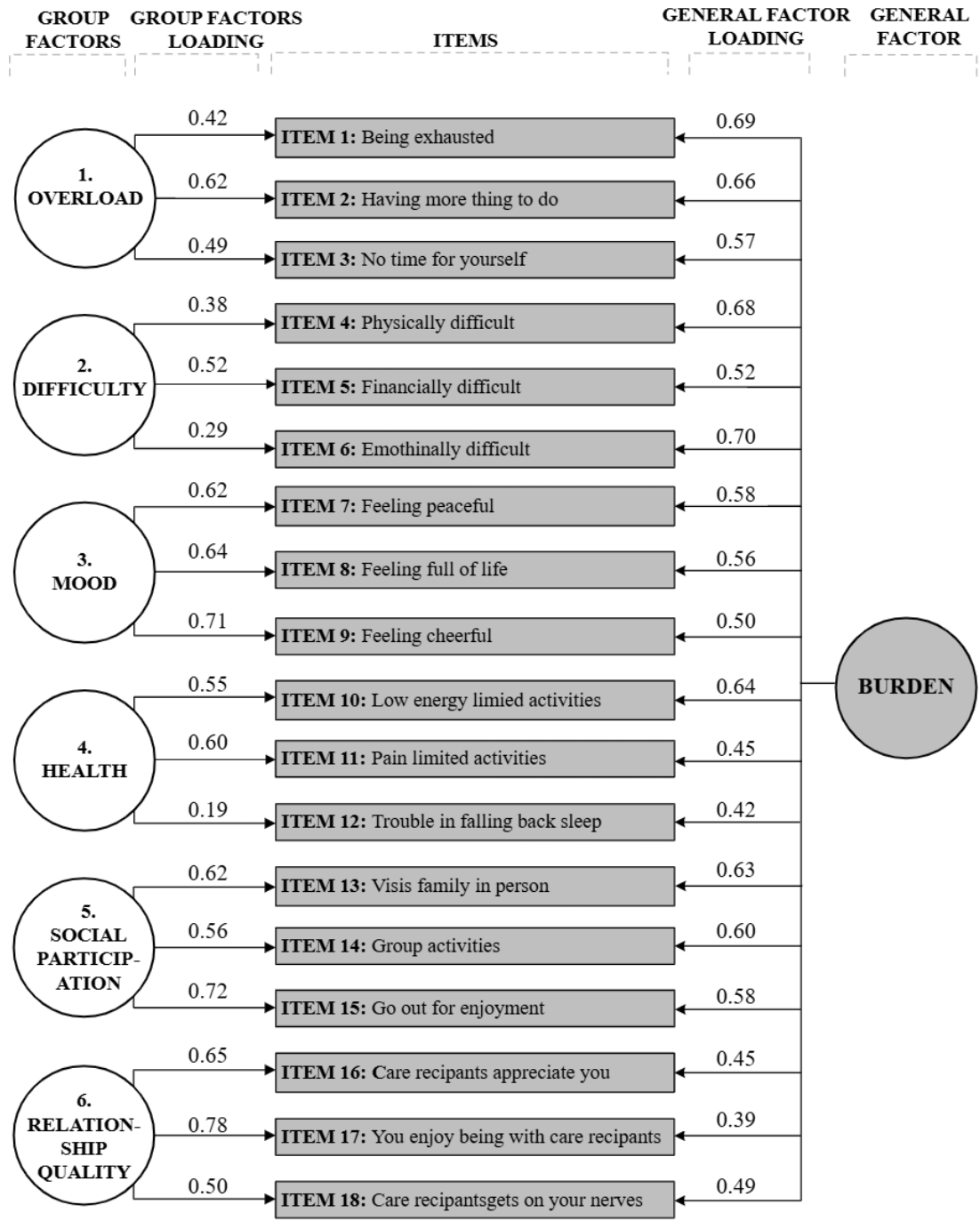

**Figure 2.** The final bifactor model for selected items. This model shows a general "BURDEN" factor and several specific group factors that directly and simultaneously predict the individual item responses. All values are standardized factor loadings.

A higher-order model was also separately determined that the six first-order factors (e.g., difficulty, social participation) loaded onto three distinct second-order factors (i.e., objective, subjective, and interpersonal), which in turn loaded onto a single, comprehensive burden factor. However, this model was used only for conceptual interpretation and not for scoring. Details are provided in the Supplementary Figure 2.

K-means clustering identified three distinct natural clusters within the scores. Based on the midpoints between cluster centers, we defined the CareBI score range as: Low Burden (0-30 score), Moderate burden (31-50 score), and High burden (51-100 score).

**Scale Validation Results**

Higher CareBI scores were a predictor of negative caregiver outcomes using both the continuous (0-100) and categorical (low, moderate, high) score representations (Table 1). For example, every one standard deviation increase in the CareBI score was associated with increased odds of a caregiver reporting frequent sleep interruption (Odds Ratio [OR] = 3.20), and feeling down (OR = 3.52). CareBI (continuous and categorical representations) also showed an increased odds of burden negatively impacting paid work, volunteer work, and caring for other family members. Moreover, higher burden scores were significantly associated with lower odds of positive experiences, such as feeling confident in one's caregiving abilities (OR = 0.73 for continuous and OR = 0.70 for categorical CareBI) and satisfaction from providing care (OR = 0.55 for continuous and OR = 0.54 for categorical CareBI). The continuous CareBI score was also a significant, albeit modest, predictor of a higher number of overnight hospital stays for the care recipient (Rate Ratio = 1.08, $p < 0.05$).

Discriminant validity of the CareBI scale was evaluated by assessing its relationship with three distinct types of NSOC questions not considered in the CareBI measure: realized supports, requested supports, and managing caregiving tasks. Realized support questions reflect assistance the caregiver receives from other family and friends in providing care. Requested support questions represent the formal services that a caregiver seeks out to address their own or the care recipient's care needs. Managing caregiving tasks questions reflect additional activities the caregiver performs to address specific care recipient needs (Table 2).

As would be expected, realized support questions were related to lower burden; for instance, care- givers who received informal help with caregiving had a 9% lower mean burden score (Ratio = 0.91) and 30% lower odds of being in a higher burden category (OR = 0.70). Conversely, requesting (or using) formal supports, an indicator of higher need, was significantly associated with higher burden scores. For example, caregivers using formal care services had a 21% higher mean burden score and were 2.24 times more likely to be in a higher burden category. Similarly, the task of finding a paid helper to assist with household chores or personal care increased the mean burden score (Ratio = 1.18) and nearly doubled the odds of a caregiver being in a higher burden category compared to those who do not seek out such help.

11**Table 1.** Criterion validity: Association of CareBI with caregivers' and care recipients' outcomes.

|  | Continuous CareBI | | Categorical CareBI | |
|---|---|---|---|---|
| **Outcome variables** | Odds Ratio | 95% CI | Odds Ratio | 95% CI |
| **Caregivers' outcomes** | | | | |
| Frequency of sleep interruption | 3.20*** | [2.80, 3.65] | 3.69*** | [3.15, 4.33] |
| Frequency of feeling upset | 3.26*** | [2.91, 3.64] | 3.66*** | [3.19, 4.21] |
| Frequency of feeling nervous | 3.16*** | [2.84, 3.51] | 3.69*** | [3.24, 4.21] |
| Frequency of feeling down | 3.52*** | [3.07, 4.04] | 4.61*** | [3.85, 5.53] |
| Frequency of feeling bored | 1.49*** | [1.37, 1.63] | 1.53*** | [1.37, 1.70] |
| Frequency of loneliness | 2.39*** | [2.17, 2.64] | 2.62*** | [2.32, 2.95] |
| Loneliness from lack of friends | 2.07*** | [2.07, 2.07] | 2.32*** | [2.04, 2.64] |
| Giving up on self-improvement | 1.58*** | [1.58, 1.58] | 1.62*** | [1.43, 1.84] |
| Feeling good about yourself | 3.16*** | [2.72, 3.67] | 3.73*** | [3.10, 4.48] |
| Frequency of uncontrollable worrying | 2.77*** | [2.47, 3.11] | 3.16*** | [2.73, 3.65] |
| Confidence in caregiving abilities | 0.73*** | [0.67, 0.80] | 0.70*** | [0.62, 0.78] |
| Increased closeness with care recipients | 0.53*** | [0.47, 0.59] | 0.49*** | [0.43, 0.56] |
| Satisfaction from providing care | 0.55*** | [0.48, 0.63] | 0.54*** | [0.45, 0.64] |
| Impact on paid work | 5.51*** | [5.46, 5.55] | 11.76*** | [11.73, 11.78] |
| Impact on volunteer work | 17.06*** | [17.03, 17.09] | 59.16*** | [10.68, 327.81] |
| Impact on caring for others | 8.56*** | [2.96, 24.79] | 7.32*** | [7.3, 7.34] |
| Significant weight loss | 1.20*** | [1.09, 1.32] | 1.29*** | [1.13, 1.46] |
| **Care recipients' outcomes** | | | | |
| Number of overnight hospital stays | 1.08* [a] | [1.00, 1.17] | 1.10* [a] | [0.99, 1.22] |
| Any overnight hospital stays | 1.17 | [0.72, 1.18] | 1.21 | [0.65, 2.24] |

*Note.* Odds Ratios (OR) reported unless otherwise indicated. CI = confidence interval.
[a] Metric reported is a Rate Ratio (RR)
* $p < 0.1$, ** $p < 0.05$, *** $p < 0.01$

Convergent validity of CareBI was confirmed by testing its association with a wide range of variables that reflect the intensity, difficulty, and scope of caregiving (Table 3). The results show a strong positive relationship between greater caregiving demands and higher CareBI scores across all tested domains. For instance, factors reflecting the intensity of care, such as the number of hours spent helping per day, and the difficulty of care (e.g., managing complex medical tasks), were both significant predictors of a higher burden score. Similarly, the scope of caregiving duties, such as providing a greater number of different care types (Ratio = 1.10) and assisting with critical tasks like banking (Ratio = 1.20), was associated with a higher mean burden.

**Table 2**. Discriminant validity: Association of the CareBI with realized and requested supports and with tasks related to managing caregiving.

| Predictors | Continuous CareBI | | Categorical CareBI | |
|---|---|---|---|---|
| | Ratio of Means | 95% CI | Odds Ratio | 95% CI |
| **Realized Supports** | | | | |
| Informal help with daily activities | 0.95* | [0.90, 1.00] | 0.86* | [0.72, 1.02] |
| Informal help with caregiving | 0.91*** | [0.86, 0.96] | 0.70*** | [0.57, 0.85] |
| **Requested supports** | | | | |
| Use of formal care services | 1.21*** | [1.120, 1.3] | 2.24*** | [1.71, 2.92] |
| Received caregiving training | 1.04 | [0.95, 1.14] | 1.48** | [1.07, 2.03] |
| **Managing caregiving tasks** | | | | |
| Assistance with finding financial help | 1.13*** | [1.04, 1.22] | 1.67*** | [1.26, 2.22] |
| Help getting mobility devices | 1.17*** | [1.12, 1.23] | 1.95*** | [1.63, 2.33] |
| Help finding paid household helper | 1.18*** | [1.10, 1.25] | 2.01*** | [1.60, 2.54] |
| Help with home safety modifications | 1.15*** | [1.09, 1.21] | 1.75*** | [1.46, 2.09] |

*Note.* CI = confidence interval.
* $p < 0.1$, ** $p < 0.05$, *** $p < 0.01$

## Discussion and Implications

In this study we developed and validated CareBI, a comprehensive single-value metric measuring the severity of caregiver burden using observational data from the NSOC. CareBI provides clinicians and clinical researchers with an unprecedented ability to understand and interpret the multi-dimensional concept of caregiver burden within observational data and aligning with prospective measures of burden assessment. We also systematically address an important gap in health service research, where validated tools, like ZBI, are incompatible with secondary, or observational data.

Our final bifactor model had strong model fit and successfully isolated a reliable general burden factor ($\omega_h = 0.74$) from the distinct influences of the specific sub-domains (e.g., "mood" or "overload"). This mathematical separation supports multiple levels of interpretation, facilitating the development of interventions targeting specific issues, such as financial difficulty or social isolation, as direct paths to mitigating the caregiver's overall burden.

**Table 3**. Convergent validity: Association of the CareBI with measures of caregiving intensity and scope.

|  | Continuous CareBI | | Categorical CareBI | |
|---|---|---|---|---|
| **Predictors** | Ratio of Means | 95% CI | Odds Ratio | 95% CI |
| **Frequency of helping:** | | | | |
|   Household chores | 1.08*** | [1.05, 1.11] | 1.42*** | [1.29, 1.55] |
|   Shopping | 1.06*** | [1.03, 1.08] | 1.30*** | [1.18, 1.42] |
|   Getting around | 1.07*** | [1.05, 1.10] | 1.33*** | [1.21, 1.45] |
|   Driving to care recipient | 1.02* | [1.00, 1.05] | 1.13** | [1.03, 1.23] |
|   Personal care | 1.08*** | [1.05, 1.11] | 1.39*** | [1.27, 1.53] |
| **Helping on a regular or varied schedule** | 1.05*** | [1.03, 1.08] | 1.27*** | [1.15, 1.39] |
| **Hours per day spending helping** | 1.14*** | [1.11, 1.17] | 1.85*** | [1.68, 2.04] |
| **Difficulty of helping:** | | | | |
|   Managing medical tasks | 1.17*** | [1.14, 1.20] | 2.00*** | [1.81, 2.21] |
|   Coordinating medical care | 1.10*** | [1.08, 1.13] | 1.60*** | [1.45, 1.76] |
|   Health insurance matters | 1.11*** | [1.08, 1.13] | 1.65*** | [1.50, 1.82] |
|   Transitioning care after a hospital stay | 1.09*** | [1.06, 1.12] | 1.44*** | [1.32, 1.58] |
| **Paying care-related expenses** | 1.08*** | [1.05, 1.11] | 1.42*** | [1.29, 1.55] |
| **Type of helping:** | | | | |
|   Prescribed medicines | 1.13*** | [1.07, 1.19] | 1.63*** | [1.35, 1.96] |
|   Banking | 1.20*** | [1.13, 1.27] | 2.03*** | [1.68, 2.47] |
|   Exercises | 1.06* | [0.99, 1.13] | 1.25** | [1.01, 1.55] |
|   Special diet | 1.10*** | [1.03, 1.17] | 1.40*** | [1.13, 1.74] |
|   Dentures | 1.16*** | [1.07, 1.25] | 1.79*** | [1.36, 2.35] |
|   Caring feet | 1.07** | [1.01, 1.14] | 1.21* | [0.98, 1.48] |
|   Shopping | 1.16*** | [1.09, 1.23] | 1.83*** | [1.50, 2.24] |
| **Number of care type** | 1.10*** | [1.07, 1.13] | 1.45*** | [1.31, 1.59] |

*Note.* CI = confidence interval.
* $p < 0.1$, ** $p < 0.05$, *** $p < 0.01$

While prior studies using the NSOC dataset either categorized burden simply as present or ab-sent without measuring its severity or analyzed its domains separately (Riffin et al. (2019); Halpern et al. (2017); Spillman et al. (2014)), CareBI provides a systematically constructed, single-value measure of burden severity. The methodology provides a reproducible template for applying quantitative approaches to complex psychosocial measurement, offering a tool to integrate the metric of caregiver well-being into the public policy and predictive modeling frameworks that inform health care decisions.

CareBI validation analyses confirmed the scale is a reliable and responsive indicator of caregiving severity, finding that burden reducing actions, such as obtaining direct caregiving help, reduces the odds of being in a higher burden category by 30%, compared to not receiving this support. Gao & Tang (2024) had similar findings in a census dataset of over 55 thousand people in Beijing, where they found that receiving meal assistance, spiritual care, and respite care reduced the likelihood of perceived emotional burden of family caregivers. Moreover, those with high monthly caregiving hours or emotional difficulty helping were more likely to seek out hospice care for support (Sullivan et al., 2022, 2023). Likewise, overnight hospitalizations of persons living with dementia is associated with caregivers reporting having financial difficulty, too much to handle, or no time for themselves (Sullivan et al., 2023).

Ham et al. (2025) found that better physical functioning in older adults was associated with lower physical difficulty for caregivers. A related finding by Sullivan et al. (2023) observed that persons living with dementia were 2.36 times more likely to have an overnight hospital stay and are 1.96 times more likely to have multiple hospitalizations when their caregivers found it physically difficult to provide care. These findings are aligned with our observation that assistance with mobility and personal care tasks is related to higher CareBI scores supporting the interpretation of CareBI as a burden severity indicator that is responsive to recipient functional needs.

The categorical CareBI representation further showed that individuals who take on the additional task of finding external services to assist with household tasks or respite needs have nearly doubled burden score than those who are not seeking help with these tasks, which may be a direct consequence of a care situation that has already become overwhelming. These findings correlate with recent analyses by Zhou & Chan (2024) who found that receiving professional assistance to find household support, nursing care for activities of daily living, and mental health needs moderated caregiver burden. Therefore, clinicians and researchers should view high CareBI scores as a red flag signaling that a caregiver is already experiencing a high degree of distress and is in urgent need of intervention.

The limitation inherent to secondary data analysis is the restriction to items available within the NSOC dataset. The inclusion of some general emotional stress items relied on attribution theory, a well-established but still theoretical assumption that caregivers would connect these feelings to their role. For model development, while some items might not be the best choice for sub-domain factors, to have orthogonal factors (i.e., uncorrelated) and a more reliable general factor, an item with lower factor loading in sub-domain was retained (item 12). To have good model fit in the complex dataset, correlation between item 6 and factor 6 was allowed. However, the study's strengths include the use of polychoric correlation to handle complex data types, strong model fit, reliable purified general factor, model interpretability, and a data-driven k-means clustering approach to establish practical low, moderate, and high burden categories. Future research should focus on applying the CareBI longitudinally to track changes in burden over time and test its application in diverse populations to guide resource allocation and support interventions for this critical group.

**Funding**
None reported.

**Conflict of Interest**
None reported.

**Data Availability**
This study used data from the National Study of Caregiving (NSOC). Access to NSOC is restricted and available only upon request through the National Health and Aging Trends Study (NHATS) website (www.nhats.org) with approval of a Data Use Agreement. The analytic methods and code used for scale development are available from the corresponding author upon reasonable request. No preregistration was required or completed for the studies reported in this manuscript.

**Acknowledgements**
National Health and Aging Trends Study. Produced and distributed by www.nhats.org with funding from the National Institute on Aging and Office of Behavioral and Social Sciences Research (OB-SSR) (grant number U01AG032947).

National Study of Caregiving. Produced and distributed by www.nhats.org with funding from the National Institute on Aging (grant numbers R01AG054004 (NSOC III) and R01AG062477 (NSOC IV)).

**References**

Al-Rawashdeh, S. Y., Lennie, T. A., & Chung, M. L. (2016). Psychometrics of the Zarit Burden Interview in caregivers of patients with heart failure. *Journal of Cardiovascular Nursing*, *31*(6), E21-E28.

Alwateer, M., Atlam, E.-S., Abd El-Raouf, M. M., Ghoneim, O. A., & Gad, I. (2024). Missing data imputation: A comprehensive review. Journal of Computer and Communi- cations, 12(11):53–75.

Bentler, P. M. (1990). Comparative fit indexes in structural models. *Psychological Bulletin*, *107*(2), 238.

Boateng, G. O., Neilands, T. B., Frongillo, E. A., Melgar-Quiñonez, H. R., & Young, S. L. (2018). Best practices for developing and validating scales for health, social, and behavioral research: a primer. *Frontiers in public health*, *6*, 149.

Brooks, M. E., Kristensen, K., Van Benthem, K. J., Magnusson, A., Berg, C. W., Nielsen, A., ... & Bolker, B. M. (2017). glmmTMB balances speed and flexibility among packages for zero-inflated generalized linear mixed modeling.

Cheung, G. W., Cooper-Thomas, H. D., Lau, R. S., & Wang, L. C. (2024). Reporting reliability, convergent and discriminant validity with structural equation modeling: A review and best-practice recommendations. *Asia Pacific Journal of Management*, *41*(2), 745-783.

Corner, S. (2009). Choosing the right type of rotation in PCA and EFA. *JALT testing & evaluation SIG newsletter*, *13*(3), 20-25.

Costello, A. B., & Osborne, J. (2005). Best practices in exploratory factor analysis: Four recommendations for getting the most from your analysis. *Practical assessment, research, and evaluation*, *10*(1).

DeVellis, R. F., & Thorpe, C. T. (2021). *Scale development: Theory and applications*. Sage publications.

Eden, J., & Schulz, R. (Eds.). (2016). Families caring for an aging America. National Academies Press.

Elmstahl, S., Malmberg, B., & Annerstedt, L. (1996). Caregiver's burden of patients 3 years after stroke assessed by a novel caregiver burden scale. *Archives of physical medicine and rehabilitation*, *77*(2), 177-182.

Forbes, Miriam K., Ashley L. Greene, Holly F. Levin-Aspenson, Ashley L. Watts, Michael Hallquist, Benjamin B. Lahey, Kristian E. Markon et al. "Three recommendations based on a comparison of the reliability and validity of the predominant models used in research

444

on the empirical structure of psychopathology." *Journal of Abnormal Psychology* 130, no. 3 (2021): 297.

Freedman, V. A., & Wolff, J. L. (2025). *The National Study of Caregiving (NSOC) Caregiver Strain Scale* (NHATS Technical Paper No. 54). Johns Hopkins Bloomberg School of Public Health. Available at https://www.nhats.org

Gao, X., & Tang, Y. (2024). Association between community elderly care services and the physical and emotional burden of family caregivers of older adults: evidence from Beijing, China. *The International Journal of Aging and Human Development*, *99*(2), 247-262.

Green, M. R., Hughes, M. C., Afrin, S., & Vernon, E. (2024). Caregiver policies in the United States: a systematic review. *Journal of Public Health Policy*, *46*(1), 22.

Griffin, J. M., Mandrekar, J. N., Vanderboom, C. E., Harmsen, W. S., Kaufman, B. G., Wild, E. M., ... & Holland, D. E. (2024). Transitional palliative care for family caregivers: outcomes from a randomized controlled trial. *Journal of pain and symptom management*, *68*(5), 456-466.

Halpern, M. T., Fiero, M. H., & Bell, M. L. (2017). Impact of caregiver activities and social supports on multidimensional caregiver burden: analyses from nationally-representative surveys of cancer patients and their caregivers. *Quality of Life Research*, *26*(6), 1587-1595.

Ham, Y., Jin, Y., Hong, I., & Park, J. H. (2025). Association between chronic illnesses in older adults and caregiver burden: a cross sectional study in the United States. *Aging & Mental Health*, 1-7.

Han, J., Kamber, M., & Mining, D. (2006). Concepts and techniques. *Morgan kaufmann*, *340*(1), 94104-103205.

Horn, J. L. (1965). A rationale and test for the number of factors in factor analysis. *Psychometrika*, *30*(2), 179-185.

Iecovich, E. (2011). Quality of relationships between care recipients and their primary caregivers and its effect on caregivers' burden and satisfaction in Israel. *Journal of Gerontological Social Work*, *54*(6), 570-591.

Kaiser, H. F. (1974). An index of factorial simplicity. *Psychometrika*, *39*(1), 31-36.

Kelley, H. H., & Michela, J. L. (1980). Attribution theory and research. *Annual review of psychology*, *31*(1), 457-501.

Kim, B., Wister, A., O'dea, E., Mitchell, B. A., Li, L., & Kadowaki, L. (2023). Roles and experiences of informal caregivers of older adults in community and healthcare system navigation: a scoping review. *BMJ open*, *13*(12), e077641.

Kuha, J. (2004). AIC and BIC: Comparisons of assumptions and performance. *Sociological methods & research*, *33*(2), 188-229.

Li, C. H. (2016). Confirmatory factor analysis with ordinal data: Comparing robust maximum likelihood and diagonally weighted least squares. *Behavior research methods*, *48*(3), 936-949.

MacQueen, J. (1967). Some methods for classification and analysis of multivariate observations. In *Proceedings of the Fifth Berkeley Symposium on Mathematical Statistics and Probability* (Vol. 1, pp. 281-297).

Marinacci, L. X., Sterling, M. R., Zheng, Z., & Wadhera, R. K. (2025). Health-Related Quality of Life of Family Caregivers in the United States, 2021–2022: A National Cross-Sectional Analysis. *Journal of General Internal Medicine*, *40*(2), 508-510.

Marsh, H. W., Hau, K. T., & Wen, Z. (2004). In search of golden rules: Comment on hypothesis-testing approaches to setting cutoff values for fit indexes and dangers in overgeneralizing Hu and Bentler's (1999) findings. *Structural equation modeling*, *11*(3), 320-341.

National Health and Aging Trends Study. (2011). Produced and distributed by www. nhats. org with funding from the National Institute on Aging (grant number NIA U01AG32947).

Reise, S. P., Moore, T. M., & Haviland, M. G. (2010). Bifactor models and rotations: Exploring the extent to which multidimensional data yield univocal scale scores. *Journal of personality assessment*, *92*(6), 544-559.

Revelle, W. (2025). *psych: Procedures for psychological, psychometric, and personality research* (Version 2.5.6) [R package]. Northwestern University. https://CRAN.R-project.org/package=psych

Riffin, C., Van Ness, P. H., Wolff, J. L., & Fried, T. (2019). Multifactorial examination of caregiver burden in a national sample of family and unpaid caregivers. *Journal of the American Geriatrics Society*, *67*(2), 277-283.

Robinson, B. C. (1983). Validation of a caregiver strain index. *Journal of gerontology*, *38*(3), 344-348.

Robinson-Lane, S. G., Johnson, F. U., Tuyisenge, M. J., Kirch, M., Christensen, L. L., Malani, P. N., ... & Koumpias, A. M. (2023). Racial and ethnic variances in preparedness for aging in place among US adults ages 50–80. *Geriatric Nursing*, *54*, 357-364.

Rosseel, Y. (2012). lavaan: An R package for structural equation modeling. *Journal of Statistical Software*, *48*(1), 1-36.

Schulz, R., & Martire, L. M. (2004). Family caregiving of persons with dementia: prevalence, health effects, and support strategies. *The American Journal of Geriatric Psychiatry*, *12*(3), 240-249.

Siminoff, L. A., Wilson-Genderson, M., Chwistek, M., & Thomson, M. (2024). The cancer caregiving burden trajectory over time: varying experiences of perceived versus objectively measured burden. *The Oncologist*, *29*(7), 629-637.

Spillman, B. C., Wolff, J., Freedman, V. A., & Kasper, J. D. (2014). Informal caregiving for older Americans: An analysis of the 2011 National Study of Caregiving. *Washington, DC: Office of the Assistant Secretary for Planning and Evaluation*.

Sullivan, S. S., Bo, W., Li, C. S., Xu, W., & Chang, Y. P. (2022). Predicting hospice transitions in dementia caregiving dyads: An exploratory machine learning approach. *Innovation in Aging*, *6*(6), igac051.

Sullivan, S. S., de Rosa, C., Li, C. S., & Chang, Y. P. (2023). Dementia caregiver burdens predict overnight hospitalization and hospice utilization. *Palliative & Supportive Care*, *21*(6), 1001-1015.

Thurstone, L. L. (1934). The vectors of mind. *Psychological review*, *41*(1), 1.

Vespa, J. E., Armstrong, D. M., & Medina, L. (2018). *Demographic turning points for the United States: Population projections for 2020 to 2060* (pp. 25-1144). Washington, DC: US Department of Commerce, Economics and Statistics Administration, US Census Bureau.

Wolff, J. L., Cornman, J. C., & Freedman, V. A. (2025). The Number Of Family Caregivers Helping Older US Adults Increased From 18 Million To 24 Million, 2011–22: Article examines growth in the number of family caregivets providing help to older adults in the US. *Health Affairs*, *44*(2), 187-195.

Zarit, S. H., Anthony, C. R., & Boutselis, M. (1987). Interventions with care givers of dementia patients: comparison of two approaches. Psychology and aging, 2(3):225.

Zhou, Y., & Chan, W. C. H. (2024). Utilization of home-based care and its buffering effects between dementia caregiving intensity and caregiver burden in China. *BMC geriatrics*, *24*(1), 913.

Zhu, M., & Ghodsi, A. (2006). Automatic dimensionality selection from the scree plot via the use of profile likelihood. *Computational Statistics & Data Analysis*, *51*(2), 918-930.

**Supplementary Material**

**Table 1**. Conceptual alignment of 44 items from the NSOC survey with the domains of the ZBI-22.

| Burden domains | Items of ZBI-22 | Pool of items |
|---|---|---|
| **Work overload** | <ul><li>Overall, how **burdened do you feel** in caring for your relative?</li><li>Do you feel your **relative is dependent upon you**?</li><li>Do you feel that your **relative asks for more help** than they need?</li><li>Do you feel that because of the time you spend with your relative that you don't have enough **time for yourself**?</li></ul> | <ul><li>When you are last helping care recipient, you are **exhausted** when you go to bed at night.</li><li>As soon as you get a routine going, care recipient **needs change**.</li><li>When you are last helping care recipient, you have **more things to do** than you can handle.</li><li>When you are last helping care recipient, you don't have **time for yourself**.</li></ul> |
| **Task difficulty** | <ul><li>Do you feel that you don't have enough **money** to care for your relative, in addition to the rest of your expenses?</li><li>Do you feel that you don't have as much **privacy** as you would like, because of your relative?</li></ul> | <ul><li>If difficult, how **physically** difficult is helping care recipient?</li><li>If difficult, how **financially** difficult is helping care recipient?</li><li>If difficult, how **emotionally** difficult is helping care recipient?</li><li>How often did helping care recipient cause your **sleep to be interrupted**?</li></ul> |
| **Mental Well-being** | <ul><li>Are you **afraid what the future** holds for your relative?</li><li>Do you feel you have **lost control of your life** since your relative's illness?</li></ul> | <ul><li>How often have you been unable to stop or control **worrying**?</li><li>How often have you **had little interest** or pleasure in doing things?</li><li>How often have you felt **nervous**, anxious, or on edge?</li><li>How often have you felt **down**, depressed, or hopeless?</li><li>How often did you feel **upset**?</li><li>I **gave up** trying to improve my life a long time ago.</li></ul> |
|  |  | - How often did you feel **bored**?
- How often did you feel **lonely**?
- I often feel lonely because I have **few close friends.**
- How often did you feel **cheerful**?
- How often did you feel calm and **peaceful**?
- How often did you feel **full of life**?
- My life has **meaning** and purpose.
- I **like my living** situation very much. |
| **Health** | - Do you feel your **health has suffered** because of your involvement with your relative? | - Would you say that in general, your **health** is excellent, very good, good, fair, or poor?
- How often did you have trouble **falling back asleep**?
- If limited, how often did your **low energy** or exhaustion limit your activities?
- If limited, how often has **pain** limited your activities? |
| **Social** | - Do you feel stressed between caring for your relative and trying to meet other **responsibilities for your family or work**?
- Do you feel that your relative currently affects your **relationship with other family members** or friends in a negative way?
- Do you feel that your **social life** has suffered because you are caring for your relative?
- Do you feel **uncomfortable about having friends over**, because of your relative? | - Did helping care recipient ever keep you from **working** for pay?
- Did helping care recipient ever keep you from **doing volunteer work**?
- Did helping care recipient ever keep you from **caring for a child** or other adult?
- If kept, how important is it to you to **visit in person** with friends or family not living with you?
- If kept, how important is it to you to **attend religious services**?
- If kept, how important is it to you to participate in **club meetings or group** activities?
- If kept, how important is it to you to **go out for enjoyment**? |
| **Self confidence** | - Do you **wish you could just leave the care of your relative** to someone else? | - Helping them has made you more **confident** about your abilities. |
| | - Do you **feel uncertain** about what to do about your relative?<br>- Do you feel that you will be **unable to take care** of your relative much longer?<br>- Do you feel you could **do a better job** in caring for your relative?<br>- Do you feel you should be **doing more** for your relative? | - Helping them has taught you to **deal with difficult** situations.<br>- Helping care recipient has given you **satisfaction** that they were well cared for.<br>- I have an **easy time adjusting** to changes.<br>- I **get over** (recover from) illness and hardship quickly.<br>- In general, **I feel confident** and good about myself. |
| Relation | - Do you feel **angry** when you are around your relative?<br>- Do you feel **strained** when you are around your relative?<br>- Do you **feel embarrassed** over your relative's behavior?<br>- Do you feel that **your relative seems to expect you** to take care of him/her, as if you were the only one he/she could depend on? | - How much does care recipient **argue** with you?<br>- How much does care recipient get on your **nerves**?<br>- How much does care recipient **appreciate** what you do for them?<br>- Helping care recipient has brought you **closer** to them.<br>- How much do you **enjoy** being with care recipient? |

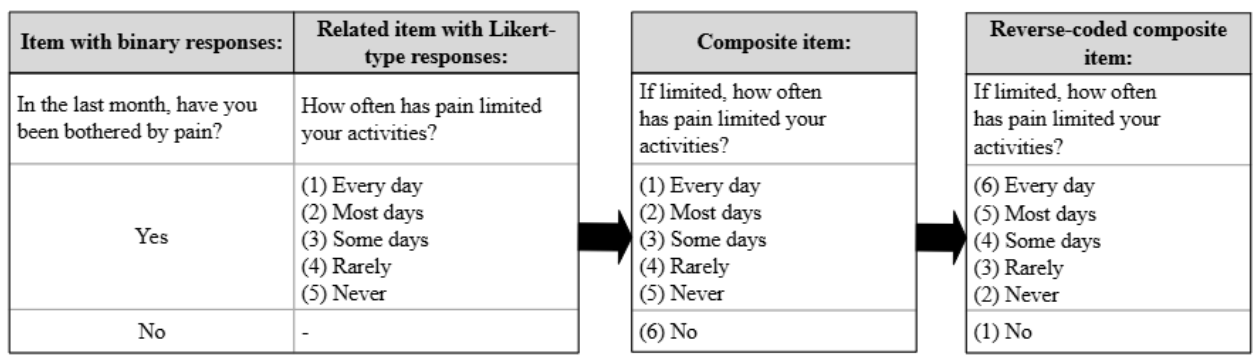

**Figure 1**. An example of a transformed item, where two related items are used to create a composite item. Since the initial composite response scale is ordered in the opposite direction of the measured distress, it is subsequently reverse-coded.

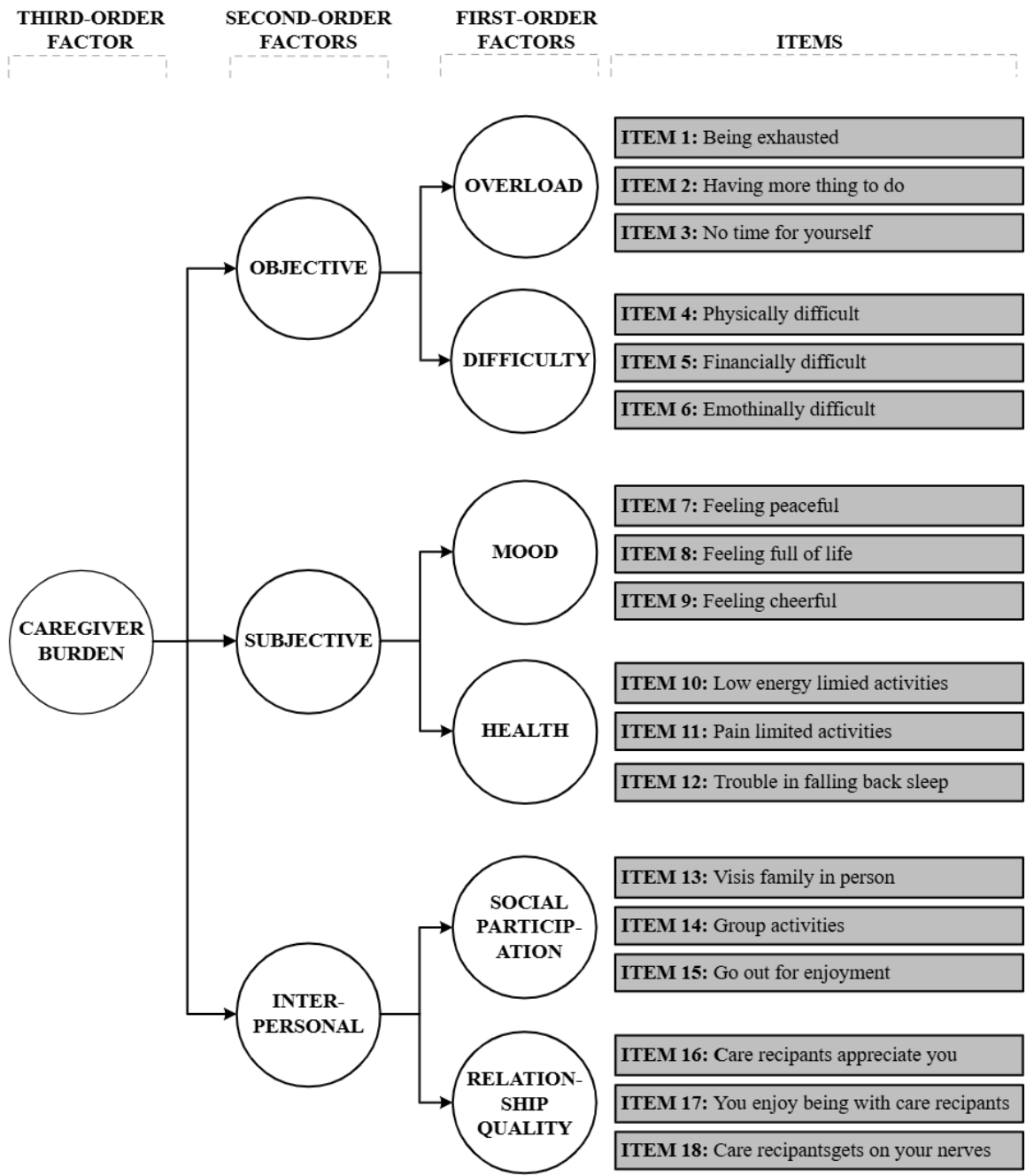

**Figure 2**. The final higher-order factor model of caregiver burden. This model shows a hierarchical structure where the general factor's influence on items is mediated indirectly through subordinate factors. This hierarchical model showed a strong fit to the data (robust CFI = 0.945, robust TLI = 0.933, robust RMSEA = 0.067).

**Table 2.** Final 18 items to estimate CareBI.

| **For the Overload domain:** |
|---|

**1)** In the last month when you were helping (care recipient), you were exhausted when you went to bed at night (cac12exhaustd).

**Answers:**

1 = Not so much
2 = Somewhat
3 = Very much

**2)** In the last month when you were helping (care recipient), you had more things to do than you could handle (cac12toomuch).

**Answers:**

1 = Not so much
2 = Somewhat
3 = Very much

**3)** In the last month when you were helping (care recipient), you did not have time for yourself (cac12notime).

**Answers:**

1 = Not so much
2 = Somewhat
3 = Very much

| **For the Difficulty domain:** |
|---|

**4)** If physically difficult helping (care recipient) has been for you (cac12diffphy), how physically difficult helping has been for you (cac12diffphlv)?

**Answers:**

1 = No, helping (care recipient) has not been physically difficult
2 = Yes, a little difficult
…
5 = Yes, very difficult

**5)** If financially difficult helping (care recipient) has been for you (cac12diffinc), how financially difficult helping has been for you (cac12diffinlv)?

**Answers:**

1 = No, helping (care recipient) has not been financially difficult
2 = Yes, a little difficult
…
5 = Yes, very difficult

**6)** If emotionally difficult helping (care recipient) has been for you (cac12diffemo), how emotionally difficult helping has been for you (cac12diffemlv)?

**Answers:**

1 = No, helping (care recipient) has not been emotionally difficult
2 = Yes, a little difficult
…
5 = Yes, very difficult

**For the Mood domain:**

**7)** In the last month, how often did you feel calm and peaceful (che12moodpcfl)?

**Answers:**

1 = Every day
2 = Most days
3 = Some days
4 = Rarely
5 = Never

**8)** In the last month, how often did you feel full of life (che12moodfull)?

**Answers:**

1 = Every day
2 = Most days
3 = Some days
4 = Rarely
5 = Never

**9)** In the last month, how often did you feel cheerful (che12moodcher)?

**Answers:**

1 = Every day
2 = Most days
3 = Some days
4 = Rarely

5 = Never

**For the Health domain:**

10) In the last month, if you had low energy or you were easily exhausted (che12low-enrgy), how often did your low energy or exhaustion limit your activities (che12enrgylmt)?

**Answers:**

1 = No, I did not have low energy or were not easily exhausted
2 = Yes, Never
3 = Yes, Rarely
4 = Yes, Some days
5 = Yes, Most days
6 = Yes, Every day

11) In the last month, if you have been bothered by pain (che12pain), how often has pain limited your activities (che12painlmt)?

**Answers:**

1 = No, I have not been bothered by pain
2 = Yes, Never
3 = Yes, Rarely
4 = Yes, Some days
5 = Yes, Most days
6 = Yes, Every day

12) In the last month, on nights when you woke up before you wanted to get up, how often did you have trouble falling back asleep?

**Answers:**

1 = Never
2 = Rarely
3 = Some nights
4 = Most nights
6 = Every night

**For the Social Participation domain:**

**13)** In the last month, if helping (care recipient) ever kept you from visiting in person with friends or family not living with you (cpp12hlpkptvs), how important is this activity to you (cpp12impvst)?

**Answers:**

1 = No, helping (care recipient) did not keep me from this activity
2 = Yes, and this activity is 'not so important' to me
3 = Yes, and this activity is 'somewhat important' to me
4 = Yes, and this activity is 'very important' to me

**14)** In the last month, if helping (care recipient) ever kept you from participating in club meetings or group activities (other than religious services) (cpp12hlpkptgr), how important is this activity to you (cpp12impgroup)?

**Answers:**

1 = No, helping (care recipient) did not keep me from this activity
2 = Yes, and this activity is 'not so important' to me
3 = Yes, and this activity is 'somewhat important' to me
4 = Yes, and this activity is 'very important' to me

**15)** In the last month, if helping (care recipient) ever kept you from going out for enjoyment (e.g., dinner, movie) (cpp12hlpkptgo), how important is this to you (cpp12impgo)?

**Answers:**

1 = No, helping (care recipient) did not keep me from this activity
2 = Yes, and this activity is 'not so important' to me
3 = Yes, and this activity is 'somewhat important' to me
4 = Yes, and this activity is 'very important' to me

**For the relationship quality domain:**

**16)** How much does (care recipient) appreciate what you do for (him/her) (cac12spapprlv)?

**Answers:**

1 = A lot
2 = Some
3 = A little
4 = Not at all

**17)** How much do you enjoy being with (care recipient) (cac12joylevel)?

**Answers:**

1 = A lot
2 = Some
3 = A little

> 4 = Not at all
>
> **18)** How much does (care recipient) get on your nerves (cac12nerveslv)?
>
> **Answers:**
>
> 1 = Not at all
> 2 = A little
> 3 = Some
> 4 = A lot

**Table 3**. Steps to estimate CareBI in practice.

> **How to apply CareBI in practice:**
>
> 1 - Select the 18 final items (see Supplementary Table 2).
>
> 2 - Create composites and reverse-code as needed so that higher values always indicate more burden (worked example in Supplementary Figure1).
>
> 3 - Fit the bifactor model (e.g., in R using *cfa* from *lavaan* package with *WLSMV*) and extract general factor scores.
>
> 4 - Extract factor scores from the fitted model; estimate the general burden factor (e.g., in R using *lavPredict* from *lavaan* package with type of latent variables "*lv*")
>
> 5 - Rescale scores to 0–100 using min–max normalization.
>
> 6 - Optional: Classify into Low (0–30), Moderate (31–50), and High (51–100) burden.